# Crystal growth and superconductivity of FeSe$_x$

S. B. Zhang, Y. P. Sun[*], X. D. Zhu, X. B. Zhu, B. S. Wang, G. Li, H. C. Lei, X. Luo, Z. R. Yang, W.H.Song and J.M. Dai

Key Laboratory of Materials Physics, Institute of Solid State Physics, Chinese Academy of Sciences, High Magnetic Field Laboratory, Chinese Academy of Sciences, Hefei 230031, China

**ABSTRACT**

In this work, crystals FeSe$_x$ have been grown by flux approach. The crystallization process is divided into two stages. First, stoichiometric polycrystal FeSe$_{0.82}$ were sintered in a solid state reaction. Then, FeSe$_x$ crystals with a size about 500μm have been successfully grown in evacuated sealed quartz tube using a NaCl/KCl flux. The products include two crystal structures of tetragon and hexagon. The electronic transport and magnetic properties measurements of FeSe$_x$ crystal exhibit a superconducting transition at about 10K.

*Corresponding author. Tel: 086-551-5592757；Fax: 086-551-5591434
**E-mail address:** ypsun@issp.ac.cn **(Y. P. Sun)**

## 1. Introduction

Recently, the discovery of 26 K superconductivity in iron arsenide $LaO_{1-x}F_xFeAs$[1] has generated a great deal of interest, and the superconductivity transition temperature $T_c$ has quickly increased to more than 40K when substituting La by other lanthanides with smaller ionic radius[2, 3, 4, 5]. It is supposed to that REOFePn (with RE = rare earth; Pn= P, As) is the first system where Fe-element plays the key role to the occurrence of superconductivity. The parent REOFePn materials possess a simple tetragonal crystal symmetry (ZrCuSiAs-type, space group P4/nmm), comprising layers of edge-sharing $FeAs_4$ tetrahedra interleaved with REF layers. These discoveries have catalyzed the search for superconducting compositions in related materials in which two-dimensional FeQ (Q = non-metal ions) slabs are also present. And the PbO-type structure $\alpha$-$FeSe_{1-x}$ has been reported that it has superconducting transition at about 8K for x = 0.12 and 0.18 [6]. Subsequent work has revealed resistivity onsets for the superconducting transition at temperature as high as 13.5K at pressure [7, 8]. FeSe has several phases: (1) a tetragonal phase $\alpha$-FeSe with PbO-structure showing a transformation from tetragonal to orthorhombic symmetry below 70K at ambient pressure [9]; (2) a NiAs-type β phase with a wide range of homogeneity showing a transformation from hexagonal to monoclinic symmetry; (3) an $FeSe_2$ phase which has orthorhombic marcasite structure. (4) Okazaki et al. have done a lot of detailed researches about $FeSe_x$ (x=8/7~4/3). Defects of iron atoms have ordered arrangements in these compounds. And new phases were found in this region. If choose the orthohexagonal and pseudo-orthohexagonal unit cells for x=8/7 and 4/3, respectively, the unit cell dimensions of the superstructure resulting from this ordering are twice as lager as that of the fundamental structure along a- and b-axes, and three times along the c-axis for x=8/7[10 - 12]. The tetragonal phase PbO-type $\alpha$-FeSe has a Fe based planar subslattice equivalent to layered iron-based oxypnictides, which has a layered crystal structure belonging to the P4/nmm space group. The crystal of $\alpha$-FeSe is composed of a stack of edge-sharing $FeSe_4$-tetrahedra layer by layer [6]. Until now, the research about $\alpha$-FeSe is mostly concentrated in polycrystal samples, and it is difficult to get a pure FeSe phase polycrystal sample using the technology reported in the past reference [6]. Therefore, a single crystal sample of FeSe is in urgently needed for the study of superconducting and structural transformation properties.

In this work, we have grown crystals of FeSe$_x$ by the flux method using NaCl/KCl as the flux in an evacuated sealed quartz tube. The structural, electronic transport and magnetic properties have been studied to characterize the property of the crystals.

## 2. Experimental procedure

Crystals of FeSe$_x$ were grown using high-temperature flux method in an evacuated sealed quartz tube. Firstly, FeSe$_{0.82}$ ploycrystal powder was prepared by the method reported in ref. 6 using high purity powders of appropriate selenium and iron stoichiometry as raw materials. The obtained FeSe$_{0.82}$ powders with several times of flux (NaCl/KCl 1/1: mol ration) were grounded and sealed into an evacuated quartz tube. The quartz tube was placed inside a vertical resistance furnace and heated to a final temperature of 850℃. A soak time of 2h was used to ensure sufficient solution of the raw materials. Then the furnace was cooled at a rate of 3℃/h to 600℃. Finally, the furnace was cooled rapidly to room temperature to avoid possible twinning. The FeSe$_x$ crystals were separated from flux by desolving the NaCl/KCl in de-ionized water.

The crystal structure of the crystals were determined by X-ray diffraction pattern (XRD) using a Philips X' pert PRO x-ray diffractometer with Cu Kα radiation at room temperature. The resistance measurement was performed by the standard four-probe method using a Quantum Design Physical Property Measurement System (PPMS) (1.8 K ≤ T ≤ 400 K, 0 T ≤ H ≤ 9 T). Magnetization measurements as a function of temperature were performed in a Quantum Design Superconducting Quantum Interference Device (SQUID) system (1.8 K ≤ T ≤ 400 K, 0 T ≤ H ≤ 5 T).

## 3. Results and discussion

### 3.1. Phase and Structure analysis

The obtained crystals are black and mirror-like in a typical size of 0.5 × 0.5 × 0.03 mm$^3$. Some flakes are rectangle, while some flakes are hexagon. Fig. 1 shows the typical photos of FeSe$_x$ crystals with two different shapes. The estimated mol ratio of Fe: Se is about 1: 1 by Energy Dispersive X-ray Spectrum (EDS), and the results have been given in Fig. 2.

Because the size of crystal FeSe$_x$ is too small, we chose several pieces of flakes of crystal FeSe$_x$ to perform the 2θ XRD pattern, and show the results in Fig. 2. The XRD shows that the samples have two crystal structures, tetragonal and hexagonal symmetry with space group of P4/nmm and P6$_3$/mmc. This result corresponds to the shapes of FeSe$_x$ crystal flakes. It is suggested that both tetragonal and hexagonal phase FeSe$_x$ crystal can be grown by this method. The calculated lattice constant is $c$ = 5.8747 Å, using the 2θ data of (00l) peak for tetragonal FeSe$_x$ crystal, which is larger than $c$ = 5.4861 Å reported before [6, 13]. And according to the 2θ data of (h00) and (00l) peak for hexagonal FeSe$_x$ crystal, the lattice constant is $a = b = 3.1168$Å, $c = 5.8839$Å.

*3.2. Superconductivity*

The resistance measurement was performed by the standard four-probe method using on PPMS. Firstly, the sample was put on a copper flake with insulating glue. Then the epoxy was used to adhere gold line (dia. 25μm) to the crystal sample as probes under a microscope. Flowing, the product has been annealed for 2h in Ar flow to volatilize the organic compound in epoxy. Finally, the product has been link to equipment by silver glue for measurement on PPMS. The resistance has an increase after annealed.

The temperature dependence of resistance in *ab* plane (2 K < $T$ < 310 K) under zero field for FeSe$_x$ crystal is plotted in Fig. 3. As it can be seen, FeSe$_x$ crystal has a metallic behavior in normal state. And the resistance has a quick drop below the onset temperature about $T_c^{onset}$ = 11.9 K, and zero resistance is attained below 3.4 K, indicating that the superconductivity was truly realized in this system. The superconducting transition width (10% - 90%) is about 8 K. The larger transition width is possibly attributed to the slight oxidation when the crystal is annealed in Ar flow. There is no obvious abnormity at ~ 100K, where the FeSe has a structural transition which is assured by before reports [6, 9]. It is suggested that the structural transition has little influence on the electronic transport properties of FeSe$_x$ crystal.

A piece of flake of FeSe$_x$ crystal is too small, so we chose several pieces of flakes for magnetic measurement. Fig. 4 shows the temperature dependence of dc magnetic susceptibility for FeSe$_x$ crystal under a 10Oe field. A sharp drop indicating the onset of superconductivity appears at ~ 10.4K is observed for zero-field-cooling (ZFC) measurements, which further confirms the existence of superconductivity. And we take note of that the field-cooling (FC) measurements has a relatively large positive magnetic susceptibility below $T_c$. It is suggested to be related to the unsuperconductive $FeSe_x$ phase in the magnetic measurement sample. And in the normal state, there are two obvious magnetic abnormity occur at 30K and 100K, and the abnormity at 105K has also been pronounced in the field-cooling measurement. This abnormity suggested that $FeSe_x$ crystal has a structural transition at about 100K, which according to the before reports [6, 9].

## 4. Conclusions

In summary, $FeSe_x$ crystal was successfully grown in evacuated sealed quartz tube using a NaCl/KCl flux. Structure analysis show that the product has two crystal structures: 1) tetragonal, P4/nmm, 2) hexagonal, $P6_3$/mmc. The results of resistivity and magnetization measurements clearly show that $FeSe_x$ crystal becomes a superconductor below ~ 10 K.

**Acknowledgements**

This work was supported by the National Key Basic Research under contract No. 2006CB601005, 2007CB925002, and the National Nature Science Foundation of China under contract No.10774146, 10774147 and Director's Fund of Hefei Institutes of Physical Science, Chinese Academy of Sciences.

**References**

[1]. Y. Kamihara, T. Watanabe, M. Hirano and H. Hosono, J. Am. Chem. Soc. 3296 (2008) 130

[2]. X. H. Chen, T. Wu, G. Wu, R. H. Liu, H. Chen and D. F. Fang, Nature, 453 (2008) 761

[3]. G. F. Chen, Z. Li, D. Wu, G. Li, W. Z. Hu, J. Dong, P. Zheng, J. L. Luo, N. L.Wang, Phys. Rev. Lett. 100 (2008) 247002.

[4]. Z. A. Ren, J. Yang, W. Lu, Y. Wei, X. L. Shen, Z. C. Li, G. C. Che, X. L. Dong, L. L. Sun, F. Zhou and Z. X. Zhao, Europhys. Lett. 82 (2008) 57002.

[5]. F. C. Hsu, J. Y. Luo, K. W. Yeh, T. K. Chen, T. W. Huang, P. M. Wu, Y. C. Lee, Y. L. Huang, Y. Y. Chu, D. C. Yan and M. K. Wu, Condmat: arXiv: 0807. 2369

[6]. Lei Fang, Huan Yang, Peng Cheng, Xiyu Zhu, Gang Mu, Hai-Hu Wen, arXiv: 0803. 3978v1

[7]. Y. Mizuguchi, F. Tomika, S. Tsuda, T. Yamaguchi, Y. Takano, arXiv: 0807. 4315

[8]. L. Li, Z. R. Yang, M. Ge, L. Pi, J. T. Xu, B. S. Wang, Y. P. Sun, Y. H. Zhang, arXiv: 0809. 0128

[9]. S. Margadonna, Y. Takabayashi, M. T. McDonald, K. Kasperkiewicz, Y. Mizuguchi, Y. Takano, A. N. Fitch, E. Suard, and K. Prassides, arXiv: 0807. 4610

[10]. A. Okazaki and K. Hirakawa, J. Phys. Soc. Japan 11 (1956) 930

[11]. A. Okazaki, J. Phys. Soc. Japan 14 (1959) 112

[12]. A. Okazaki, J. Phys. Soc. Japan 16 (1961) 1162

**Figure Captions**：

Fig. 1 The Photos of FeSe$_x$ crystals with different shapes.

Fig. 2 Energy Dispersive X-ray Spectrum (EDS) of FeSe$_x$ crystals with different phases.

Fig. 3 XRD pattern on several pieces of flakes crystal FeSe$_x$.

Fig. 4 Temperature dependence of resistance of FeSe$_x$ crystal at zero field. The inset is the magnification plot of resistance in low temperature region.

Fig. 5 Temperature dependence of magnetic susceptibility $\chi(T)$ for FeSe$_x$ crystals at 10Oe. The inset is the magnification plot of $\chi(T)$ in low temperature region.

**Figures**

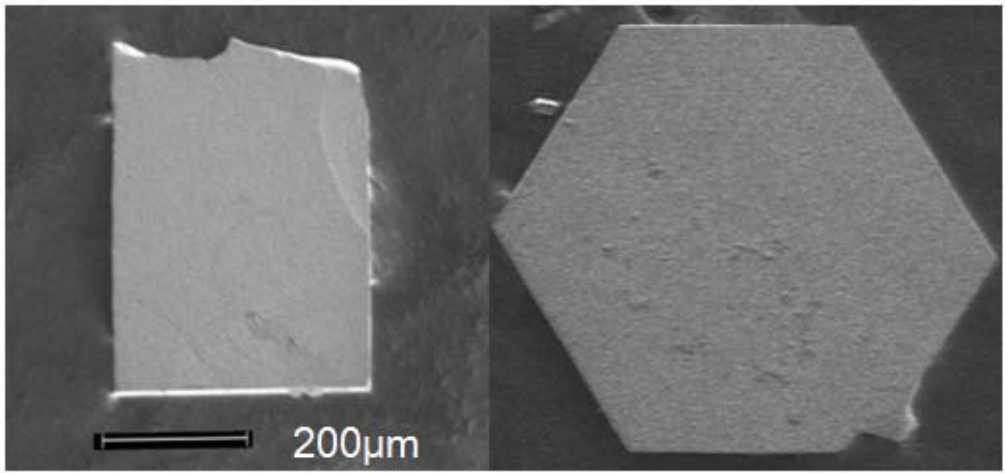

Fig. 1 S. B. Zhang et al.

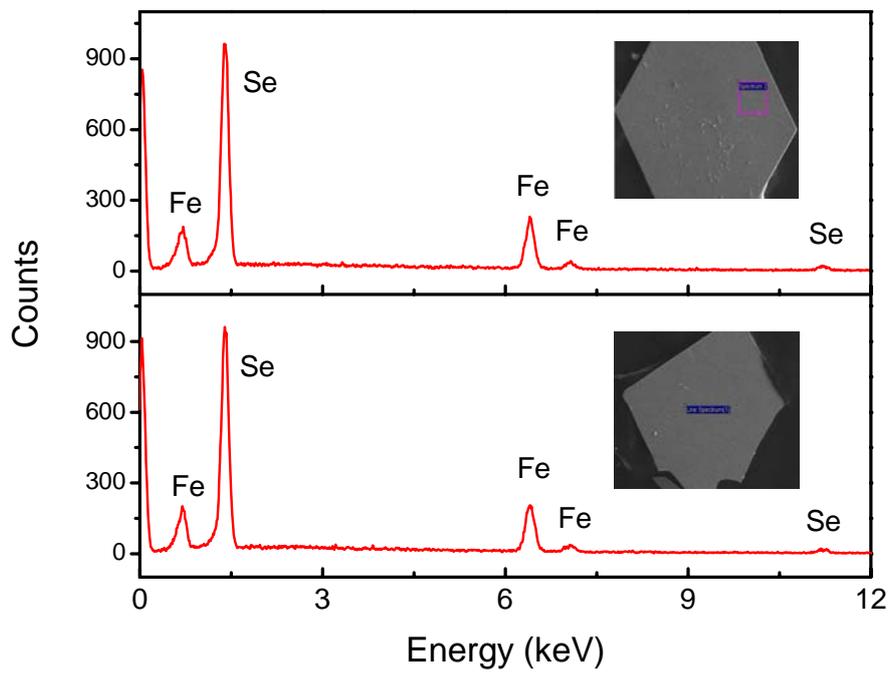

Fig. 2 S. B. Zhang et al.

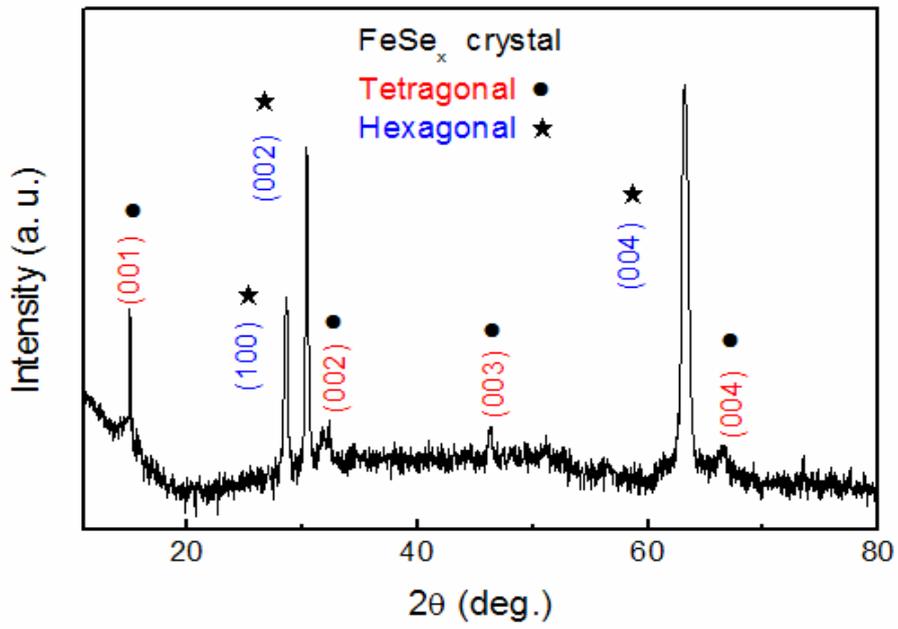

Fig. 3 S. B. Zhang et al.

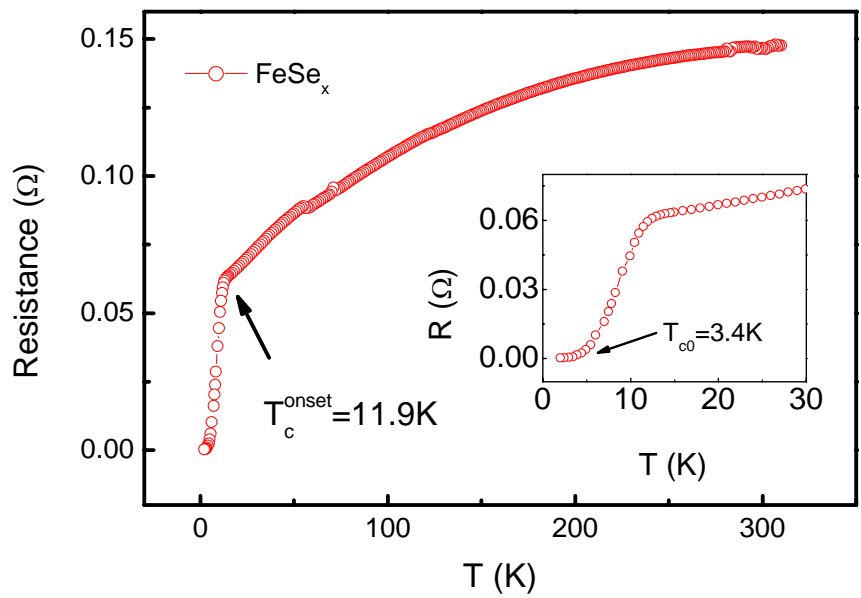

Fig. 4 S. B. Zhang et al.

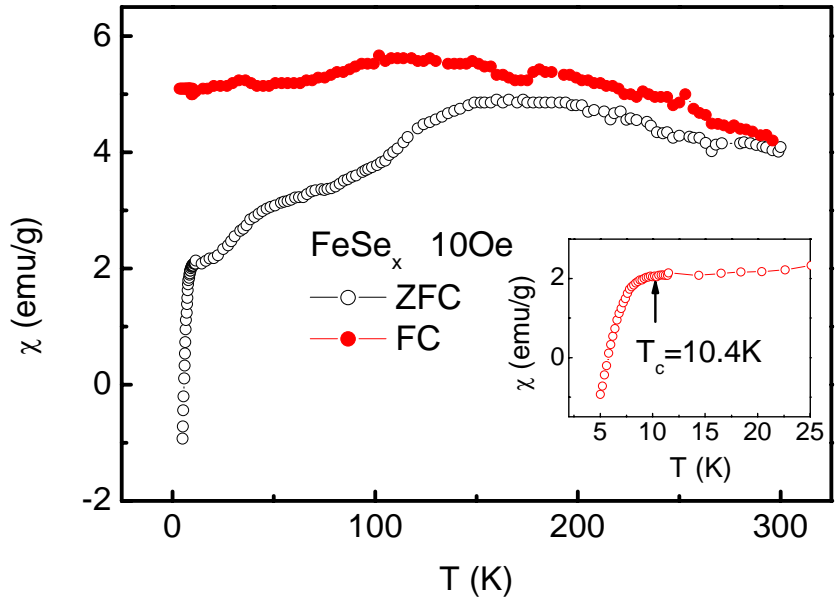

Fig. 5 S. B. Zhang et al.